\documentclass[%
 aip,
 apl,
 sd,%
 onecolumn,
 amsmath,amssymb,
preprint,%
]{revtex4-1}

\usepackage{graphicx}
\usepackage{dcolumn}
\usepackage{bm}

\begin{document}


\title {Ring-shaped polariton lasing in pillar microcavities}

\author{V.K.~Kalevich}
\email{kalevich@solid.ioffe.ru}

\author{M.M.~Afanasiev}%
\author{V.A.~Lukoshkin}
\author{K.V.~Kavokin}
\affiliation{Spin Optics Laboratory, State University of Saint-Petersburg, 1, Ulianovskaya,  198504, St-Petersburg, Russia 
}%
\affiliation{A.F. Ioffe Physico-Technical Institute, Russian Academy of Sciences, 26, Politechnicheskaya, 194021, St-Petersburg, Russia
}%

\author{S.I.~Tsintzos}
\affiliation{IESL-FORTH, P.O. Box 1527, 71110 Heraklion, Crete,
Greece
}%

\author{P.G.~Savvidis}
\affiliation{IESL-FORTH, P.O. Box 1527, 71110 Heraklion, Crete,
Greece
}%
\affiliation{Department of Materials Science and Technology,
University of Crete, Greece
}%

\author{A.V.~Kavokin}
\affiliation{Spin Optics Laboratory, State University of Saint-Petersburg, 1, Ulianovskaya,
198504, St-Petersburg, Russia 
}%
\affiliation{Physics and Astronomy School, University of
Southampton, Highfield, Southampton, SO171BJ, UK}


\begin{abstract}
Optically generated exciton-polaritons in cylindric semiconductor
pillar microcavity with embedded GaAs/AlGaAs quantum wells
demonstrate a clear polariton lasing regime. When exciting in the
center of the pillar we detect a ring-shaped emission, where the
peak of intensity can be separated from the excitation spot by
more than 10 micrometers. The spatial coherence of the ring
emission is verified by interferometry measurements. These
observations are interpreted by drift of the exciton polariton
condensate away from the excitation spot due to its repulsion from
the exciton reservoir and by its spatial confinement by the pillar
boundary.
\end{abstract}

\maketitle

Polariton lasers are devices capable of emitting spontaneously a
coherent and monochromatic light\cite{Imam1996}. They are based on
condensation of mixed light-matter quasiparticles --
exciton-polaritons -- in a single quantum state in a semiconductor
microcavity due to their bosonic nature. Polariton lasers with
optical\cite{Chris2007} and electrical
injection\cite{Schn2013,Bhatt2013} have been realized in planar
and pillar\cite{Bloch2008,Bloch2010} microcavities based on
various semiconductor materials.

Exciton-polaritons repel each other due to their excitonic
component\cite{Yamamoto1999,Ciuti2000,Kavokin2010}. In the case of
optical excitation of a planar microcavity by a sharply focused
light beam, this leads to radial drift of
polaritons\cite{Wouters2008,Vulkano2013} and formation of a
polariton condensate at some distance from the exitation spot,
unlimitedly expanding farther away\cite{Baumberg2012}. The lateral
confinement of polaritons in micropillars\cite{Ferr2011} or
microwires\cite{Wertz2010} helps shaping polariton condensates and
may lead to their patterning, which would affect the beam-shape of
the emitted light. As an example, in square pillars the emission
is located at corners\cite{Ferr2011}.

Here we report formation of ring-shaped exciton-polariton
condensates in cylindric GaAs/AlGaAs pillar microcavities under
nonresonant optical pumping in the center of the pillar by a
sharply focused continuous-wave (cw) laser. We have discovered a
ring-shaped pattern of spatially coherent emission. The diameter
of the emitting ring strongly exceeds the diameter of the
excitation spot (2\,$\mu$m). It can be as large as 30\,$\mu$m in
the pillar of 40\,$\mu$m diameter.

Changing the intensity of non-resonant optical excitation, we
clearly observe a threshold to polariton lasing which is
manifested by the increase of the intensity of emission by several
orders of magnitude and by significant narrowing of the emission
line. The distribution of polaritons in the reciprocal space
develops a sharp peak near \emph{k}\,=\,0. The spatial coherence
builds up, which is evidenced by interferometry measurements. All
these findings confirm the onset of a polariton lasing regime with
a peculiar ring-shaped pattern of emission.

\begin{figure*}
\includegraphics[width=17.0cm,angle=0]{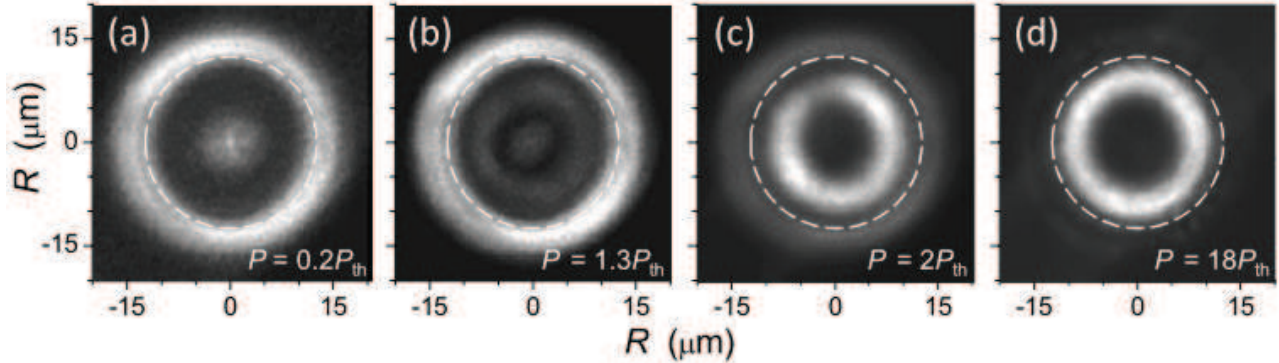}
\caption{\label{Rings} Real space images of a 25\,$\mu$m pillar
measured at pumping power $P/P_{\rm th}=0.2$ (a), 1.3 (b), 2 (c),
and 18 (d), where $P_{\rm th}=2.2$\,mW. Here $R$ is the distance
from the pillar center. White dashed circumference shows the
pillar edge. $T=3.5$\,K. The brightness in each panel is
normalized to its maximum over the image. }
\end{figure*}

We studied the set of cylindric pillars which were produced by
etching the MBE-grown planar $5\lambda/2$ Al$_{0.3}$Ga$_{0.7}$As
microcavity with top and bottom distributed Bragg reflectors
(DBRs) consisting of 32 and 35 pairs of
AlAs/Al$_{0.15}$Ga$_{0.85}$As, respectively, and having the
quality factor $Q>16000$. Four sets of three 10\,nm
Al$_{0.3}$Ga$_{0.7}$As/GaAs quantum wells (QWs) are placed at
antinodes of the cavity electric field in order to maximize the
exciton-photon coupling strength\cite{Savv2012}. A wedge in the
cavity thickness permits variation of the detuning energy
$\delta=E_{C}-E_{X}$, where $E_{C}$ and $E_{X}$ are energies of
the cavity mode and of the heavy-hole exciton at zero in-plane
wavevector (\emph{k}\,=\,0). The samples under study were placed
into the helium-flow cryostat and kept at $T=3.5$\,K. The
photoluminescence (PL) from the pillars was collected after
non-resonant excitation by a cw Ti:sapphire laser in a local
minimum of the DBR reflectivity ($\approx110$\,meV above $E_{X}$).
The laser beam was focused to a 2\,$\mu$m spot at the pillar
center by a microscope objective (focus length = 4\,mm, numerical
aperture =\,0.42). The same objective was used to collect PL. Real
space images as well as the \emph{k}-space images of pillars were
projected on the entrance slit of a 50~cm-monochromator and after
spectral dispersion were recorded by a CCD-camera. When taking
real space images, the grating of the monochromator was set to the
zeroth order of diffraction, and the width of the entrance slit
was set to 3\,mm. All experiments were performed at normal
incidence of the excitation beam on the sample surface. To
suppress the excitation laser light scattered from the pillar
surface, a cut-off interference filter was installed in front of
the entrance slit.

We studied pillars with the diameters of 16, 20, 25, 30, and
40\,$\mu$m. Since the main obtained results are qualitatively the
same for all those pillars, we present here the data for a
25\,$\mu$m one only.

Figure~\ref{Rings} shows real space images of a 25\,$\mu$m pillar
measured at different excitation powers. At the lowest pumping,
the intensity pattern in the image is a ring with a small spot in
the center (Fig.~\ref{Rings}(a)). The inner diameter of this ring
coincides with that of the pillar top shown by the dashed
circumference in Fig.~\ref{Rings}. We will show below that this
"border" ring is the exciton radiation escaping from the side
surface of the pillar. The central spot in Fig.~\ref{Rings}(a) is
mainly due to the scattered light of the excitation laser. With
the increase of pumping intensity, the second ("inner") ring of
smaller diameter appears and it becomes dominant under strong
pumping, as seen in Figs.~\ref{Rings}(b)--\ref{Rings}(d).
According to our estimations, in the strong pumping regime
($P/P_{\rm th}>3$, where $P_{\rm th}=2.2$\,mW is the polariton
laser threshold, as will be shown below) the integrated intensity
of emission of the inner ring exceeds the emission integrated over
the remaining area of the pillar by two orders of magnitude. Note
that the diameter of the inner ring slightly increases with the
pump power, not exceeding the diameter of the pillar.

\begin{figure}
\includegraphics[width=7.5cm,angle=0]{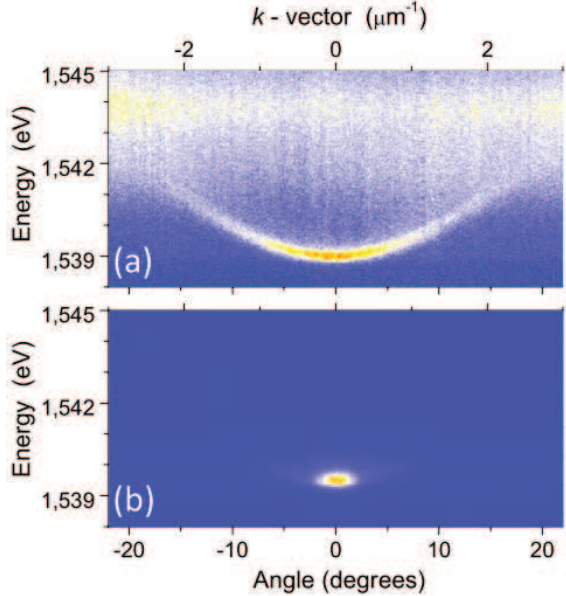}
\caption{\label{Disp} (color online). \emph{k}-space image of a
25\,$\mu$m pillar normalized to its maximum brightness at (a)
$P/P_{\rm th}=0.45$ and (b) $P/P_{\rm th}=4.5$.}
\end{figure}

To reveal the origin of the inner and border rings, we analysed
\emph{k}-space images and spatially resolved PL spectra of the
pillar at different pumping intensities.

\begin{figure*}
\includegraphics[width=17.0cm,angle=0]{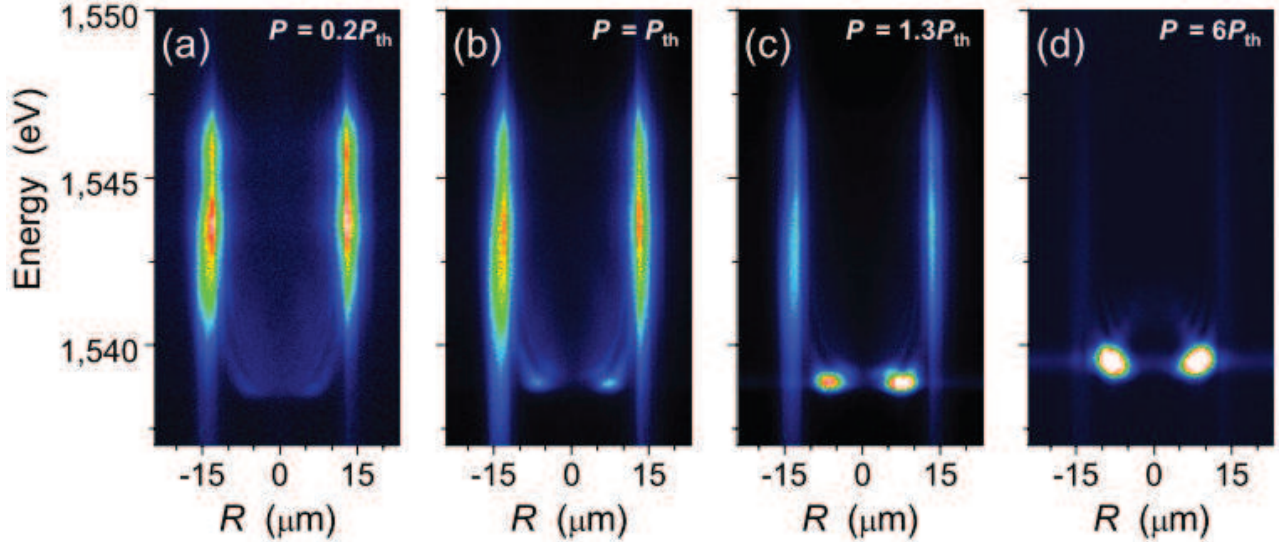}
\caption{\label{SS} (color online). Spatially and spectrally
resolved emission spectra of a 25\,$\mu$m pillar at pumping power
$P/P_{\rm th}$ equal to (a)~0.2, (b)~1, (c)~1.3, and (d)~6.
Intensity scales are different for different panels: pump power
dependences of the peak exciton and polariton intensities are
shown in Fig.~\ref{Nonl}(a). }
\end{figure*}

The \emph{k}-space images measured by the angle-resolved
PL\cite{Weisbuch1994} at low and high excitation power are shown
in Figs.~\ref{Disp}(a) and \ref{Disp}(b), respectively. At low
pumping, the polariton dispersion exhibits a minimum at 1.539\,eV
and $k=0$. We observe also a broad emission line at 1.544\,eV. It
is interpreted as the PL of heavy-hole excitons escaping from the
side surface of the pillar due to scattering at the imperfections
of this surface, as it was already found in
Ref.\,[\cite{Bloch2008}]. Simulations of the polariton and exciton
dispersions in the model of three coupled oscillators (not shown
here) allow us to conclude that in this particular pillar the
optical mode is negatively detuned from the heavy-hole exciton,
$\delta=- 2$\,meV. The increase of the excitation power results in
an abrupt narrowing of the distribution of polariton emission over
energy and angle, down to fractions of meV and a couple of angular
degrees, at pump intensity exceeding the polariton lasing
threshold, as demonstrated in Fig.~\ref{Disp}(b). This is a
signature of polariton lasing, already observed in several
microcavity systems under non-resonant
excitation\cite{Deng2002,Kasprzak2006,Balili2007}.

The diameter of the real space image of the pillar on the entrance
slit was about 2\,mm. Using a much smaller width of the entrance
slit (100\,$\mu$m), we could spectrally and spatially resolve the
photoluminescence along the diameter of the pillar.
Figure~\ref{SS} shows the spectral-spatial images taken at four
different excitation powers. At the lowest pumping, the spectrally
broad emission from the border of the pillar dominates
(Fig.~\ref{SS}(a)). This broad line can only be attributed to
emission by heavy-hole excitons. It is peaked at 1.544\,eV, which
coincides with the exciton feature observed in the \emph{k}-space
image in Fig.~\ref{Disp}(a), detected at the same excitation
power. At smaller distances \emph{R} from the pillar center, we
also observe a weak polariton emission at lower energies. The edge
of this emission spectrum at 1.539~eV is well below the exciton
resonance and coincides with the bottom of the low polariton
branch in Fig.~\ref{Disp}(a).

\begin{figure}
\includegraphics[width=7.5cm,angle=0]{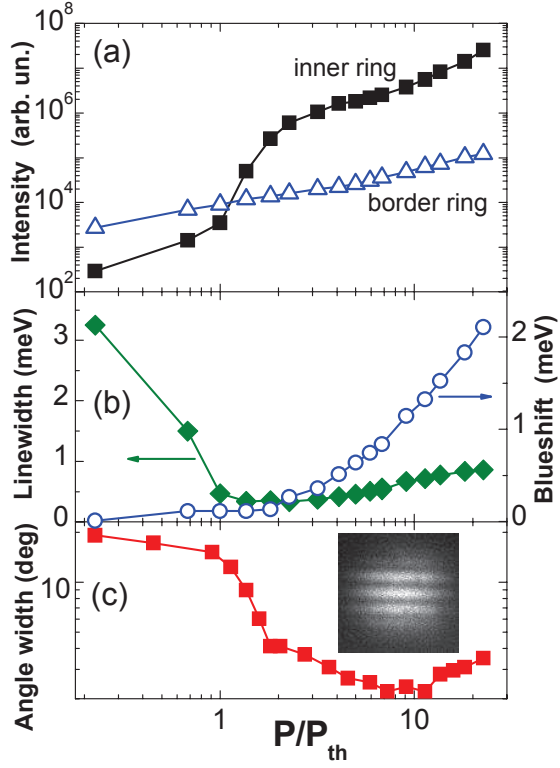}
\caption{\label{Nonl} (color online). (a) Emission intensity of
polaritons (squares) and of excitons (triangles). (b) Spectral
linewidth (diamonds) and blueshift (circles) of polariton
emission. (c) Full width at half-maximum of the angular
distribution of polariton emission. $T = 3.5$\,K. Inset in panel
(c): fringe pattern emerging when light beams from two spots in
Fig. \ref{SS}(d) are superimposed. }
\end{figure}

The polariton emission from the inner part of the pillar increases
with the pumping power. At $P\approx P_{\rm th}$ it starts
concentrating in two small spots at $|\emph{R}|\approx 7$\,$\mu$m
and the energy of 1.539\,eV, as shown in Fig.~\ref{SS}(b). Under
further increase of pumping, these two localised spots of
polariton emission become dominating over the incoherent exciton
emission (Figs. \ref{SS}(c)--\ref{SS}(d)). The distance between
these spots coincides with the inner ring diameter in Fig.
\ref{Rings}, thereby indicating that these spots originate from
the inner ring radiation. Figure \ref{Nonl}(a) shows the peak
intensity of the PL signal from one of these spots and the peak
intensity of the border emission as functions of the pumping
intensity. The border PL intensity linearly increases with the
pumping intensity (triangles in Fig. \ref{Nonl}(a)), that confirms
its excitonic origin. On the contrary, the inner-ring emission
(polariton emission) increases superlinearly (squares in Fig.
\ref{Nonl}(a)): it increases by two orders of magnitude with the
pumping power changing by only a factor of 2, from $P=P_{\rm th}$
to $P=2P_{\rm th}$. The superlinear increase of polariton emission
at $P_{\rm th}\leq P\leq 2P_{\rm th}$ is characteristic of the
formation of a polariton condensate by stimulated scattering. The
condensate formation is also confirmed by a strong spectral
(squares in Fig. \ref{Nonl}(b)) and angular (Fig. \ref{Nonl}(c))
narrowing of the polariton emission with the pumping power
increase up to $P\approx2P_{\rm th}$. A sharp, down to two
degrees, narrowing of the angular distribution of the polariton
emission at $P>>P_{\rm th}$ means that the light beam emitted by
the polariton laser has the shape of a tube directed along the
pillar axis.

Formation of the polariton condensate in the shape of a ring,
whose diameter greatly exceeds the size of the excitation spot, is
caused, in our opinion, by interaction of the polaritons with the
exciton cloud formed from electron-hole pairs created by the
nonresonant pump in the pillar center. Because of a slow loss of
energy by excitons via emission of acoustic phonons, and of a long
exciton lifetime (a considerable part of excitons under
nonresonant pumping forms outside the light cone and cannot
recombine radiatively), a large population of excitons builds up.
Since excitons have short diffusion length\cite{SSE1996}, they
accumulate in the vicinity of the excitation spot. Their repulsive
exchange interaction with polaritons creates for the latter a
potential hill in the center of the pillar\cite{Ferr2011}.
Together with the infinitely high potential barrier at the pillar
border, this results in the formation of a localizing potential
having the shape of a circular groove, which serves as a trap
facilitating polariton condensation. With increase of the pump
power and, therefore, of the concentration of excitons, the
potential near the excitation spot rises, and the diameter of the
condensate ring should become somewhat larger. Indeed, as seen in
Fig.\,\ref{SS}, the ring diameter increases from 13.6 to
16.6\,$\mu$m with the growth of $P$ from $P_{\rm th}$ to $6P_{\rm
th}$. The inner ring diameter up to 30\,$\mu$m was observed in the
40\,$\mu$m pillar.

This result can be compared with the observation by Christmann
\emph{et al}.\cite{Baumberg2012} on a planar cavity. They have
observed a radially symmetric condensate of polaritons,
accelerating outwards from the excitation spot, with a round dark
circle in the center. The size of the condensate was much larger
than the radius of the central dark circle, and it did not have a
pronounced outer boundary. In our case, the outer boundary of the
condensate is defined by the pillar edge, which is why it takes
the shape of a high-contrast, thin ring.

The polariton emission in Fig.~\ref{SS} demonstrates a blue shift,
as shown in detail in Fig.~\ref{Nonl}(b) (circles). The blue shift
may originate either from the polariton interaction with excitons
or from polariton-polariton repulsion within the condensate. When,
as in our case, the pillar diameter is much larger than the size
of the excitation spot, localization regions of the polariton
condensate and of the exciton reservoir are spatially separated.
In this case, the blue shift weakly depends on the
exciton-polariton interaction and is mainly determined by the
polariton-polariton interaction in the condensate\cite{Ferr2011}.

The inset in Fig.~\ref{Nonl}(c) shows the Young interferometry
image measured by superimposing the light beams emitted by two
spots in Fig.~\ref{SS} at $P/P_{\rm th}=6$. A clear fringe pattern
is observed beyond the polariton lasing threshold while it is
absent at $P<P_{\rm th}$. Since these two spots in Fig.~\ref{SS}
are situated at the opposite ends of the ring diameter (the arc
length between the spots is about 26\,$\mu$m), this observation
confirms the buildup of a spatial coherence in the whole ring
polariton condensate.

In this Letter we concentrated on the lowest energy polariton
state, which provides the strongest emission intensity. Weaker
emission of polaritons with higher energies, seen in
Fig.~\ref{SS}, will be considered elsewhere.

To conclude, a ring-shaped polariton condensate is formed in a
cylindric-pillar microcavity under non-resonant optical pumping at
the center of the pillar. The diameter of the ring is much larger
than the excitation spot and may be controlled by the pumping
intensity.

This work was partly supported by the Russian Ministry of
Education and Science (contract No. 11.G34.31.0067). P.S. and S.T.
acknowledge support from the Russian exchange grant EU FP7 IRSES
"POLATER" and P.S. is indebted to Greek GSRT program "ARISTEIA"
(1978) and FP7 ITN "INDEX" (289968) projects for financial
support.

\end{document}